\documentclass[%
 reprint,
 amsmath,amssymb,
 aps,
prb,
]{revtex4-2}

\usepackage{graphicx}
\usepackage{dcolumn}
\usepackage{bm}
\usepackage[colorlinks=true,linkcolor=blue,anchorcolor=blue, citecolor=blue,urlcolor=blue]{hyperref}
\begin{document}

\preprint{APS/123-QED}

\title{Spin excitations in the heavily overdoped monolayer graphene superconductor: An analog to the cuprates}

\author{Wei-Jie Lin$^1$}
\thanks{These two authors contributed equally to this work.}
\author{W. LiMing$^1$}
\thanks{These two authors contributed equally to this work.}
\author{Tao Zhou$^{1,2}$}
 \email{Corresponding author: tzhou@scnu.edu.cn}

\affiliation{$^1$Guangdong Provincial Key Laboratory of Quantum Engineering and Quantum Materials, Department of Physics, School of Physics and Telecommunication Engineering, South China Normal University, Guangzhou 510006, China\\
$^2$Guangdong-Hong Kong Joint Laboratory of Quantum Matter, Frontier Research Institute for Physics, South China Normal University, Guangzhou 510006, China
}

\date{\today}

\begin{abstract}
Recently it was reported experimentally that the monolayer graphene can be heavily overdoped to beyond the Van Hove regime.
We study theoretically the possible superconductivity and the corresponding spin excitations of the monolayer graphene in this doping region.
A static spin-density-wave state is favorable due to the nested Fermi surface as the Fermi level is doped to the Van Hove singularity point. Superconductivity may be realized upon further doping. The spin excitations in the superconducting state are studied theoretically based on the random phase approximation. The overall features are qualitatively the same with those in cuprate superconductors due to similar Fermi surface topologies of these two families of materials.
Thus we have proposed an exciting possibility, namely, the recently realized beyond-van-Hove graphene
is a cuprate analog and can become a novel platform to study the unconventional superconductivity.
\end{abstract}

\maketitle

\section{introduction}

High-T$_c$ superconductivity in the family of cuprate materials has been studied intensively for
more than thirty years, while so far its mechanism remains puzzling~\cite{RevModPhys.78.17}. This has motivated the great effort to seek for cuprate analogs. Previously, several possible candidate superconducting families have been proposed, including the iron-based superconductors~\cite{RevModPhys.83.1589}
 and the recently discovered nickelate superconductor~\cite{Li2019}.

The realization of superconductivity in graphene-based materials has been paid considerable attention since the first production of graphene in 2004~\cite{RevModPhys.81.109}. In the past several years, evidence of superconductivity has indeed been reported in several graphene-based materials with different methods~\cite{PhysRevLett.111.246805,Ichinokura2016,PMID:26979564,DiBernardo2017,Ludbrook11795,Cao2018,Park2021}. Especially, it was reported that the twisted bilayer graphene will exhibit flat bands near the Fermi energy, leading to the correlated parent insulating
states~\cite{Caoy2018}. The superconductivity emerges upon doping of this parent state. As a result, the twisted bilayer graphene has become a novel platform to study unconventional superconductivity and has attracted tremendous interest~\cite{Cao2018}.

The band structure of the monolayer graphene has saddle points at the $M$ point of the Brillouin zone. The quasiparticle dispersion near this point is flat leading to the divergent density of states according the Van Hove singularity (VHS) scenario. Of particular interest is to tune the Fermi energy to the VHS point upon doping (the VHS filling)~\cite{PhysRevLett.104.136803,PhysRevB.100.035445,PhysRevB.100.121407}. Then the Fermi surface is perfectly nested leading to a spin-density-wave (SDW) instability~\cite{Li_2012,PhysRevB.84.125404,PhysRevB.85.035414,PhysRevB.86.020507}. On the other hand, the VHS near the Fermi level may provide an effective attractive potential then the superconducting pairing is also favored~\cite{PhysRevB.81.085431,PhysRevLett.104.136803,PhysRevB.85.035414,PhysRevB.86.020507,Nandkishore2012,PhysRevB.89.144501,Xiao_2016,Xu_2019}.
Previously the competition between the SDW order and the superconducting pairing has been studied theoretically~\cite{PhysRevLett.104.136803,PhysRevB.85.035414,PhysRevB.86.020507}. Such competition is of interest and is indeed similar to that in the Cuprate compound.

Very recently, it was reported that the graphene doping technique reaches a new level, namely, the Fermi level is doped to beyond the VHS regime for the first time~\cite{PhysRevLett.125.176403}. The electronic structure of the monolayer graphene at this doping region is similar to that of the twisted graphene material~\cite{Cao2018,Caoy2018,Park2021}. At the VHS filling, the magnetic order is induced by the flat band at the Fermi energy. The static magnetic order is suppressed and the superconducting state may emerge upon further doping.
Therefore, the recently realized monolayer graphene beyond the VHS regime may become another platform to study the unconventional superconductivity.

For cuprate superconducting materials, it is generally believed that the spin fluctuation may play a fundamental role and mediate the superconducting pairing. Experimentally, the momentum and energy dependence of spin excitations are obtained directly through the inelastic neutron scattering (INS) experiments~\cite{ROSSATMIGNOD199186,PhysRevB.46.5561,PhysRevLett.86.1610,PhysRevLett.83.608,Bourges1234,Hayden2004,Vignolle2007,PhysRevLett.90.137004,PhysRevLett.110.177002,PhysRevB.98.060502,PhysRevLett.125.117002}.
Theoretically, the INS experimental results can be compared through
exploring the imaginary part of the dynamical spin susceptibility~\cite{PhysRevB.68.224503,PhysRevLett.82.2915,PhysRevB.96.014515,PhysRevB.76.094510,PhysRevB.70.212512,PhysRevB.78.140509,PhysRevB.78.020514,PhysRevB.79.235207,Zhou2020}.
For the graphene-based material, it has been proposed theoretically that the spin fluctuation may account for the superconductivity~\cite{Xiao_2016,PhysRevB.103.L041103}, while so far there is still no experimental evidence for the spin excitations. It is understandable because the Fermi surface pockets are generally small for the graphene at the low doping region~\cite{Li2021}.
The Fermi surface becomes a large pocket when the doping density exceeds the VHS filling.
Based on the Fermi surface nesting scenario~\cite{PhysRevB.68.224503,PhysRevLett.82.2915,PhysRevB.96.014515,PhysRevB.76.094510,PhysRevB.70.212512,PhysRevB.78.140509,PhysRevB.78.020514,PhysRevB.79.235207,Zhou2020}, the graphene material at this doping region should also have strong spin fluctuation.
Since the heavily overdoped monolayer graphene beyond the VHS filling has just been realized experimentally~\cite{PhysRevLett.125.176403},
we expect that the spin excitations in this material may be tested by experiments later.
Therefore, now it is timely and of importance to study the spin fluctuation in the heavily overdoped graphene superconductor and compare the results with those in the high-T$_c$ superconducting materials.

In this paper, we study theoretically the dynamic spin susceptibility of the mono-layer graphene material at the doping level beyond the VHS filling based on the random phase approximation (RPA). We consider a typical $d+id$ pairing state, consistent with previous theoretical predictions~\cite{PhysRevB.81.085431,PhysRevLett.104.136803,PhysRevB.85.035414,PhysRevB.86.020507,Nandkishore2012,PhysRevB.89.144501,Xiao_2016,Xu_2019}. The resonant spin excitation is revealed. The overall features of spin excitations are qualitatively the same with those in high-T$_c$ superconductors, indicating that these two families of materials have qualitatively similar fermiologies.
 Therefore, we propose that the heavily overdoped monolayer graphene materials are indeed analogous to the cuprate materials and may become a novel platform to study the unconventional superconductivity.

The rest of the paper is organized as follows.
In Sec. II, we introduce
the model and present the relevant formalism. In Sec. III, we
report numerical calculations and discuss the obtained
results. Finally, we present the brief summary in Sec. IV.

\section{Model and formalism}

We start from the Hamiltonian including the bare superconducting Hamiltonian and an onsite repulsive interaction, expressed as,
\begin{equation}
H=H_{0}+H_{\rm int}.
\end{equation}

$H_{0}$ includes the hopping term, the chemical potential term, and the superconducting pairing term. Considering the superconducting pairing between two nearest-neighbor sites, this term can be written as:
\begin{align}
H_0=&-t\sum_{\langle {\bf ij}\rangle,\sigma}(c_{{\bf i}\sigma}^{\dagger}c_{{\bf j}\sigma}+{\rm h.c.})+
\sum_{\langle {\bf ij}\rangle}(\Delta_{\bf ij}c_{{\bf i}\uparrow}^{\dagger}c_{{\bf j}\downarrow}^{\dagger}
+{\rm h.c.})\nonumber\\&-\mu\sum_{{\bf i},\sigma}
c_{{\bf i}\sigma}^{\dagger}c_{{\bf i}\sigma},
\end{align}
where ${\bf j}$ is the nearest-neighbor site of the site ${\bf i}$ with ${\bf j}={\bf i}+{\bf e_\alpha}$. Here each site ${\bf i}$ has three nearest-neighbor sites with ${\bf e_1}=\frac{1}{2}(\sqrt{3},1)$, ${\bf e_2}=\frac{1}{2}(-\sqrt{3},1)$ and ${\bf e_3}=(0,-1)$.
$\Delta_{{\bf i}{\bf j}}$ represents the superconducting pairing between the two nearest-neighbor site, with $\Delta_{{\bf i}, {\bf i}+{\bf e_j}}=\Delta_{\rm j}=\Delta_0e^{i\phi_j}$. The superconducting pairing symmetry is determined by the parity and the phase $\phi_{j}$~\cite{PhysRevB.77.235420}. For the $d+id$ pairing symmetry, generally the phase changes $2\theta_0$ when the vector ${\bf e_j}$ rotates for $\theta_0$. Therefore, in the present work, we consider $\Delta_{1,2,3}=\Delta_0 e^{-i4\pi/3}$, $\Delta_0 e^{-i8\pi/3}$, and $\Delta_0$.
In addition, the even parity with $\Delta_{\bf ij}=\Delta_{\bf ji}$ is considered.

$H_{\rm int}$ is the on-site interaction term, expressed as
\begin{align}
H_{\rm int}=U\sum_{\bf i}n_{{\bf i}\uparrow}n_{{\bf i}\downarrow}.
\end{align}

In the momentum space, the bare Hamiltonian $H_0$ can be rewritten as $H_0=\sum_{\bf k}\Psi_{\bf k}^\dagger H_{\bf k} \Psi_{\bf k}$. The vector $\Psi^\dagger_{\bf k}$ is expressed as,
\begin{equation}
\Psi^\dagger_{\bf k}=(c^\dagger_{1{\bf k}\uparrow},c^\dagger_{2{\bf k}\uparrow},c_{1,{-\bf k}\downarrow},c_{2,{-\bf k}\downarrow}),
\end{equation}
with $1$ and $2$ being the sub-lattice indices for the graphene lattice.

The $4\times4$ matrix $H_{\bf k}$ is expressed as,
\begin{equation}
H_{\mathbf{k}}
=\left(
\begin{array}{ccccc}
 -\mu & \gamma(\mathbf{k}) & 0 & \Delta(\mathbf{k})\\
 \gamma^*(\mathbf{k}) & -\mu & \Delta(-\mathbf{k}) & 0\\
 0 & \Delta^*(-\mathbf{k}) & \mu & -\gamma^*(-\mathbf{k})\\
 \Delta^*(\mathbf{k}) & 0 & -\gamma(-\mathbf{k}) & \mu\\
\end{array}
\right),
\end{equation}
with $\gamma(\mathbf{k}) = -t\sum_\mathrm{j}e^{i\mathbf{k}\cdot \mathbf{e_j}}$ and
$\Delta(\mathbf{k})=\sum_\mathrm{j}\Delta_\mathrm{j} e^{i\mathbf{k}\cdot \mathbf{e_j}}$.

The spin susceptibility can be expressed as a $4\times 4$ matrix with the elements being calculated through the spin-spin correlation function~\cite{PhysRevB.79.235207,PhysRevB.96.014515,PhysRevLett.111.066804},
\begin{equation}
\chi_{l_3,l_4}^{l_1,l_2}({\bf q},i\omega_n)=\int^\beta_0 d\tau \langle T_\tau S^{l_1,l_2+}_{\bf q}(\tau)S^{l_3,l_4-}_{\bf -q}(0)\rangle e^{i\omega_n},
\end{equation}
with the spin operator being expressed as $S^{l_1,l_2+}_{\bf q}=\sum_{\bf k}c^\dagger_{l_1{\bf k}\uparrow}c_{l_2,{{\bf k}+{\bf q}}\downarrow}$. $l_i=1,2$ is the sublattice index.
When $l_1=l_2$ and $l_3=l_4$, the operators $S^{\pm}_{\bf q}$ are corresponding to the physical spin density operators.

Without the interaction term ($H_{\rm int}=0$), the bare spin susceptibility in the superconducting state
includes both the normal and the anomalous terms,
\begin{small}
\begin{align}
\chi_{l_3,l_4(0)}^{l_1,l_2}(\mathbf{q},\omega)=\frac{1}{N}\sum_\mathbf{k}\sum_{\alpha,\beta=1}^4
[\xi_{l_4}^\alpha(\mathbf{k})\xi_{l_1}^{\alpha,*}(\mathbf{k})\xi_{l_2}^\beta(\mathbf{k}+\mathbf{q})\xi_{l_3}^{\beta,*}(\mathbf{k}+\mathbf{q}) \nonumber\\
+\xi_{l_4}^\alpha(\mathbf{k})\xi_{l_2+2}^{\alpha,*}(\mathbf{k})\xi_{l_3}^{\beta,*}(\mathbf{k}+\mathbf{q})\xi_{l_1+2}^{\beta}(\mathbf{k}+\mathbf{q})]\frac{f(E_{\mathbf{k}+\mathbf{q}}^\beta)-f(E_{\mathbf{k}}^\alpha)}
{\omega+E_\mathbf{k}^\alpha-E_{\mathbf{k}+\mathbf{q}}^\beta+i\eta}.
\end{align}
\end{small}
$E^\alpha_{\bf k}$ and $\xi^\alpha(\mathbf{k})$ are the eigenvalue and the eigenvector of the Hamiltonian matrix $H_{\mathbf{k}}$.
$f(x)$ is the Fermi distribution function.

The renormalized spin susceptibility $\chi({\bf q},\omega)$ can be obtained through the RPA, given by
\begin{align}
\hat{\chi}(\mathbf{q},\omega)=[\hat{I}-\hat{\chi}_{0}(\mathbf{q},\omega)\hat{U}]^{-1}\hat{\chi}_{0}(\mathbf{q},\omega),
\end{align}
where $\hat{I}$ is the $4\times4$ identity matrix. With onsite interaction being considered, the nonzero elements of the $\hat{U}$ matrix include
 $U_{l_3,l_4}^{l_1,l_2}=U$ for $l_1=l_2=l_3=l_4$.
The physical spin susceptibility can be obtained through the sum of the elements of $\hat{\chi}$ with $l_1=l_2$ and $l_3=l_4$.

\begin{figure}
\includegraphics[width=3.1in]{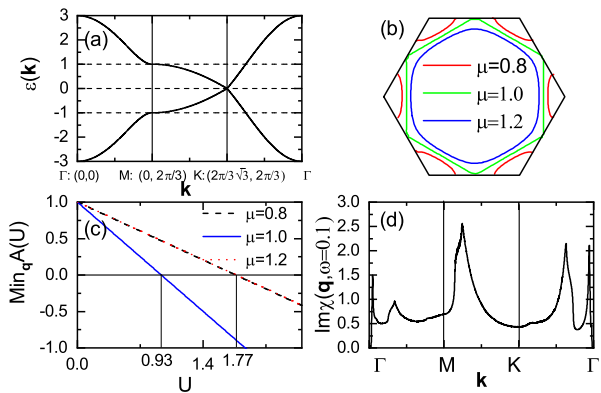}
\caption{\label{fig:1} (Color online) (a) The normal state energy bands along the highly
symmetrical lines in the Brillouin zone. (b) The normal state Fermi surfaces with different chemical potentials.
 (c) The minimum value of the RPA factor as a function of the interaction $U$ with $\omega=0$. (d) The imaginary part of the renormalized normal state spin susceptibility as a function of the momentum with $\omega = 0.1$ and $\mu=1.2$.}
\end{figure}

\section{results and discussion}

The normal state energy bands and Fermi surfaces [obtained by setting $\Delta_0=0$ in Eq.(4)] are displayed in Figs.~1(a) and 1(b), respectively. As is seen, at the $M$ point [${\bf Q_M}=(0,2\pi/3)$], the quasiparticle dispersion is flat at the energy $E=1.0$, leading to the VHS at this energy. When the chemical potential is below the VHS filling ($\mid\mu\mid<1$), the normal state Fermi surface contains several small disconnected pockets. At the VHS filling with $\mu=1$, the Fermi surface is perfectly nested. The topology of the Fermi surface changes, namely, the Fermi surface becomes one large pocket. Beyond the VHS filling ($\mu>1$), the Fermi surface keeps to be a large pocket centered at the $\Gamma=(0,0)$ point.

As discussed previously~\cite{Li_2012,PhysRevB.84.125404,PhysRevB.85.035414,PhysRevB.86.020507}, the nested Fermi surface will lead to the SDW instability. With the RPA framework, the SDW instability can be explored through the RPA factor $A({\bf q},\omega)$, with
\begin{equation}
A({\bf q},\omega)=\mathrm{det}\mid \hat{I}-\hat{\chi}_{0}(\mathbf{q},\omega)\hat{U}\mid.
\end{equation}
At the zero energy the imaginary part of $A({\bf q},\omega)$ is zero. If its real part
at a certain wave-vector ${\bf q}$ is negative, then the magnetic instability occurs. In this case the RPA method cannot be used directly and a static SDW order should be induced to describe the system.
The minimum value of the RPA factor with $\omega=0$ as a function of the onsite interaction $U$ is plotted in Fig.~1(c).
At the VHS filling $(\mu=1.0)$, the minimum value of $A(U)$ is less than zero as the interaction $U$ is larger than a critical value $U_c$ ($U_c=0.93$), indicating the signal for the SDW instability. As the chemical potential deviates from the VHS filling, this critical interaction $U_c$ increases significantly. In the following presented results,
we study the spin excitations beyond the VHS filling with
$\mu=1.2$ and $U=1.6$. The normal state Fermi surface at this chemical potential is a large pocket, similar to that of cuprate materials~\cite{RevModPhys.78.17}.
The static SDW order disappears with these parameters and the RPA technique is effective to study the spin fluctuation.
The imaginary part of the renormalized normal state spin susceptibility as a function of the momentum with $\omega=0.1$ is presented in Fig.~1(d).
As is seen, the maximum spin excitation occurs at an incommensurate momentum $(\delta,2\pi/3)$ near the $M$ point.

\begin{figure}
\includegraphics[width=3.1in]{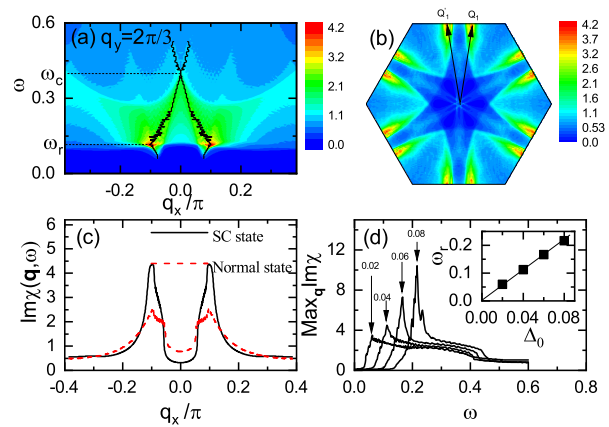}
\caption{\label{fig:2} (Color online) (a)The imaginary part of the spin susceptibility as functions of the energy and the momentum along the line
$q_y=2\pi/3$ with $\Delta_0=0.04$. (b) The imaginary part of the spin susceptibility as a function of the momentum with $\omega=\omega_r=0.12$. (c) The imaginary parts of the spin susceptibility as a function of the momentum in the superconducting state and the normal state with $\omega=0.12$. (d) The maximum value of Im$\chi({\bf q},\omega)$ in the Brillouin zone as a function of the energy with different gap magnitudes. Inset: The resonant energy $\omega_r$ as a function of the gap magnitude $\Delta_0$.}
\end{figure}

We now study the spin excitations in the superconducting state. The intensity plot of the imaginary part of the spin susceptibility (Im$\chi$) as functions of the energy and the momentum along the line $q_y=2\pi/3$ with $\Delta_0=0.04$ is presented in Fig.~2(a). Here two typical energies, i.e., $\omega_r\approx 0.12$ and $\omega_c\approx 0.4$, are revealed and indicated in Fig.~2(a). At the energy $\omega_r$, Im$\chi$ reaches its maximum value, at an incommensurate momentum
with ${\bf q}=(\pm\delta,2\pi/3)$. $\delta\approx 0.1\pi$ is the incommensurability. As the energy increases, the incommensurability decreases. At the energy $\omega_c$, the spin excitation is commensurate with the maximum value emerging at the momentum ${\bf Q_M}=(0,2\pi/3)$. As the energy increases to be larger than $\omega_c$, the spin excitation becomes incommensurate again. The dispersion of the maximum spin excitations in the whole momentum and energy space has a hourglass shape.

Let us study in more detail the spin excitation at the energy $\omega_r$. The intensity plot of the imaginary part of the spin susceptibility in the whole Brillouin zone at the energy $\omega=\omega_r$ is displayed in Fig.~2(b). As is seen, the spin excitation has six-fold symmetry with the maximum excitation emerging at the incommensurate momentums ${\bf Q_1}$ and ${\bf Q_1^\prime}$ (or their symmetrical momentums). The two dimensional cut of Im$\chi$ along the line $q_y=2\pi/3$ as a function of $q_x$ is replotted in Fig.~2(c). The imaginary part of the spin susceptibility in the normal state with $\omega=0.12$ is also plotted in Fig.~2(c). As is seen, the intensity of the incommensurate peak at the energy $\omega_r$ in the superconducting state is significantly stronger than that in the normal state.
The enhanced spin excitation in the superconducting state indicates the signal of the spin resonance mode. Previously, the resonant spin excitation has attracted broad interest in various unconventional superconductors~\cite{ROSSATMIGNOD199186,PhysRevB.46.5561,PhysRevLett.86.1610,PhysRevLett.82.2915,PhysRevB.96.014515,PhysRevB.76.094510,PhysRevB.70.212512,PhysRevB.78.140509,PhysRevB.78.020514,Zhou2020}. The resonant excitation will not necessarily appear based on the RPA freamwork and it depends strongly on the band structure and the pairing symmetry. Therefore, it has been widely used to resolve the pairing symmetry of an unconventional superconductor~\cite{PhysRevB.70.212512,PhysRevB.78.140509,PhysRevB.78.020514,Zhou2020}.
In the mean time, it has been verified that the resonant energy is proportional to the gap magnitude or the superconducting transition temperature T$_c$~\cite{PhysRevLett.86.1610,PhysRevB.76.094510}, so that it is believed to be intimately related to the superconductivity. The maximum spin excitations in the whole momentum space as a function of the energy with different gap magnitudes (from $\Delta_0=0.02$ to $\Delta_0=0.08$) are plotted in Fig.~2(d). The intensity of the maximum value of Im$\chi$ increases significantly as $\Delta_0$ increases.
The possible resonant energy $\omega_r$ can be obtained from Fig.~2(d) through the position of maximum Im$\chi$.
The energy $\omega_r$ as a function of the gap magnitude is presented in the inset of Fig.~2(d). As is seen, here $\omega_r$ is proportional to $\Delta_0$, being in accord with the main character of the spin resonance. These features verify that the resonant spin excitation indeed exists. We have also checked numerically that for the present band structure the spin resonance mode only exist for the typical $d+id$ pairing symmetry.
There are no resonant spin excitations
if another different pairing symmetry is considered. Therefore,
such resonant behavior may be used to identify the pairing symmetry and the unconventional superconductivity in
this material.

We would like to compare the above numerical results of the spin excitation with those in high-T$_c$ cuprate superconductors.
In cuprate superconductors, the spin excitations are material dependent. For the La-based material (e.g., La$_{2-x}$Sr$_x$CuO$_4$), two energy scales are revealed experimentally~\cite{Vignolle2007}. The maximum spin excitation appears at a lower energy about $18$ meV and an incommensurate momentum.
At a higher energy (about $50$ meV), the spin excitation is commensurate. For Y-based high-T$_c$ superconducting material (e.g., YBa$_2$Cu$_3$O$_{7-x}$), a resonant spin excitation at the energy about 40 meV is revealed~\cite{ROSSATMIGNOD199186,PhysRevB.46.5561}. The spin excitation at this energy is commensurate.
 Here obviously, the spin excitations with two energy scales are analogous to those in La-based high-T$_c$ superconducting materials.
 Moreover, we have checked numerically that if a rather strong gap magnitude is considered ($\Delta_0\geq0.2 t$), the spin excitation at the resonant energy will become very strong and appears at the commensurate momentum ${\bf Q_M}$. In this case the results are similar to those of Y-based superconductors.

Another important result revealed in Fig.~2(a) is the hourglass dispersion of the spin excitation, with a downward dispersion below the energy $\omega_c$ and an upward dispersion above the energy $\omega_c$. Previously, such hourglass shape dispersions are also reported in the hole-doped cuprate materials~\cite{Vignolle2007,PhysRevLett.83.608,Bourges1234,Hayden2004}, while it was reported that
there is
no low energy downward dispersion in the electron-doped cuprate material~\cite{PhysRevLett.90.137004,PhysRevB.68.224503}.
For iron-based superconductors, so far the dispersion of spin excitations is still an important and open issue and the results may also depend on the materials.
It was reported that the low energy spin excitations are commensurate for iron-based superconductors~\cite{PhysRevLett.110.177002,PhysRevB.98.060502}.
However, very recently, it was also reported that the hourglass dispersion is observed in a typical iron-based superconductor KCa$_2$Fe$_4$As$_4$F$_2$~\cite{PhysRevLett.125.117002}.
Actually, the low energy dispersion of the spin excitation depends strongly on the Fermi surface and the superconducting gap.
Here the similarity of the dispersion for spin fluctuations between the heavily overdoped monolayer graphene and the hole-doped cuprate superconducting materials is of interest. This indicates that the Fermi surface topology of the heavily overdoped graphene
is indeed analogous to that of the hole-doped cuprate materials.

\begin{figure}
\includegraphics[width=2.6in]{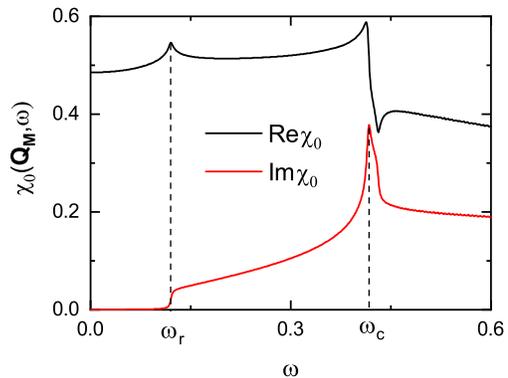}
\caption{\label{fig:3} (Color online) The real and imaginary parts of the bare spin susceptibility with $\Delta_0=0.04$.}
\end{figure}

The above features of spin excitations can be understood well based on the RPA framework and the topology of the Fermi surface. The general origin of the spin resonance in an unconventional superconductor has been studied intensively~\cite{PhysRevLett.82.2915,PhysRevB.96.014515,PhysRevB.76.094510,PhysRevB.70.212512,Zhou2020}. The renormalized spin susceptibility includes two parts of contribution, namely, the bare spin susceptibility $\chi_0({\bf q},\omega)$ and the RPA factor $A({\bf q},\omega)$. The real and imaginary parts of the bare spin susceptibility are displayed in Fig.~3. As is seen, the imaginary part of the bare spin susceptibility is nearly zero due to the presence of the spin gap. At the edge of the spin gap Im$\chi_0$ increases rapidly, then Re$\chi_0$ has a peak structure due to the Kramers-Kronig relation. As a result, the real part of the RPA factor reaches the minimum value at this energy. Thus the imaginary part of the renormalized spin susceptibility in the superconducting state will be enhanced at this energy, leading to the resonant spin excitation at this energy. At higher energies, the spin excitation is mainly determined by the topology of the Fermi surface. The bare spin excitation has the maximum value at the energy $\omega_c$, as indicated in Fig.~3. This can be understood further from the nesting of the energy contour.

\begin{figure}
\includegraphics[width=2.6in]{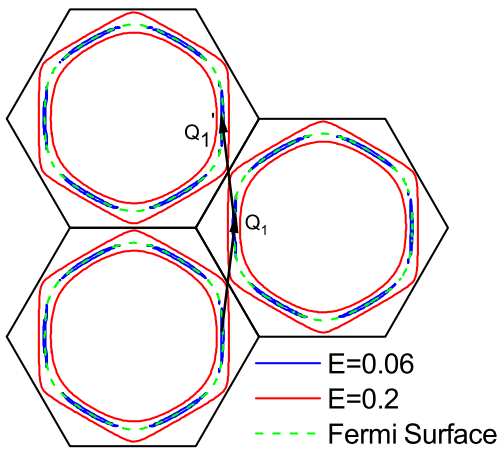}
\caption{(Color online) The constant energy contours with $E({\bf k})=0.06$ and $E({\bf k})=0.2$ in the first and the extended Brillouin zones. The (green) dashed line is the normal state Fermi surface. The arrows indicate the nesting vectors for the contour $E({\bf k})=0.06$.
}
\end{figure}

The bare spin susceptibility is mainly contributed by the particle-hole excitation. We have
 Im$\chi_0({\bf q},\omega)\propto \sum_{\bf k}\delta[\omega-\Omega({\bf q},{\bf k})]$ with $\Omega({\bf q},{\bf k})=E({\bf k})+E({\bf k}+{\bf q})$. Generally, to explain the spin excitation at a certain energy $\omega_0$, one needs to study the scattering between the energy contour $E=\omega_0/2$~\cite{PhysRevLett.82.2915}. The contour plots of the energy contours with $E=\omega_r/2=0.06$ and $E=\omega_c/2=0.2$ in the first and extended Brillouin zones are presented in Fig.~4. As is seen, the contour $E=0.06$ contains a flat piece, indicating the nesting feature. The best nesting vectors are indicated in Fig.~4, namely, ${\bf Q_1}$ and ${\bf Q_1^\prime}$, well consistent with the numerical results shown in Fig.~2. As the energy increases, the size of the energy contour becomes larger. Naturally the incommensurability will decrease. As the energy is much greater than $\Delta_0$, the superconducting pairing term plays a minor role. As a result, the energy contour $E=0.2$ almost coincides with the normal state Fermi surfaces of $\mu=1.0$ and $\mu=1.4$. Since the normal state Fermi surface with $\mu=1.0$ is perfectly nested with the nesting vector ${\bf q}={\bf Q_M}$. As a result, the spin excitation at the energy $\omega=0.4$ is commensurate and its maximum value appears at the momentum ${\bf Q_M}$. As the energy increases further, the nesting condition of the energy contour is broken and the spin excitation will become incommensurate again.

\section{summary}

At last, we summarize the similarity between the possible superconductivity in the heavily overdoped graphene material with that in cuprates. Firstly, the general phase diagram may be similar. For the graphene material, at the VHS filling, the Fermi surface is nested, leading to the SDW instability. Beyond the VHS filling, the static SDW order is suppressed and the superconductivity may emerge. For the cuprates, at the half-filling, the system is in the antiferromagnetic state. The antiferromagnetic order is suppressed and the superconductivity is realized upon doping. Secondly, although so far no consensus has been reached about the mechanism of superconductivity in cuprates, while many believe that the spin fluctuation should play an essential role. Here for the heavily overdoped graphene material, the strong spin fluctuation should exist due to the nesting of the Fermi surface. Such a strong fluctuation could account for possible unconventional superconductivity in this material. Thirdly, as presented in our present work, the overall features of spin excitations in the superconducting graphene material are qualitatively similar with those of cuprate superconductors, including the resonant spin excitation at a certain energy and the hourglass dispersion. Here the spin excitations are consistent with the topology of the Fermi surface. Our results clearly indicate that the Fermi surface topologies of these two families are similar.
Therefore, the heavily overdoped monolayer graphene material may indeed be a potential candidate to be the cuprate analog.

In conclusion, we have studied theoretically spin excitations in the superconducting state of the heavily overdoped monolayer graphene material beyond the VHS filling. Two typical energies are revealed. At a lower energy, a resonant spin excitation is revealed at an incommensurate momentum. At a higher energy, the spin excitation is commensurate. The dispersion of the spin excitation has a hourglass shape.
The overall features are qualitatively the same with those in superconducting cuprate materials. We propose that the mono-layer graphene material beyond the VHS filling is a candidate to become the cuprate analog. The spin resonance mode may be detected by later experiments and our results can be used to identify the unconventional superconductivity in this material.

\begin{acknowledgments}
This work was supported by the NSFC (Grant No. 12074130), the Natural Science Foundation of Guangdong Province (Grant No. 2021A1515012340), and Science and Technology Program of Guangzhou (Grant No. 2019050001).
\end{acknowledgments}

%

\end{document}